# Ferroelectric and anti-ferroelectric coupling in superlattices of paraelectric perovskites at room temperature

Hans M. Christen*, Eliot D. Specht[†], Sherwood S. Silliman[‡§] and K. S. Harshavardhan[‡]

*Oak Ridge National Laboratory, Condensed Matter Sciences Division, Oak Ridge, TN 37831-6056
[†]Oak Ridge National Laboratory, Metals and Ceramics Division, Oak Ridge, TN 37831-6118
[‡]Neocera, Inc., 10000 Virginia Manor Road, Beltsville, MD 20705
[§]currently at Intel Massachusetts Inc., Hudson MA 01749

Results from dielectric and structural measurements on epitaxial $SrTiO_3/BaZrO_3$ superlattices reveal properties that cannot be explained simply in terms of those measured on single films of the constituent materials. For large superlattice periodicities (20/20 and 38/38 structures, i.e. samples in which each $SrTiO_3$ and $BaZrO_3$ layer are 20 or 38 unit cells thick, respectively), the capacitance-voltage curves indicate room-temperature ferroelectricity. For smaller periodicities (7/7 and 15/15), anti-ferroelectric-type behavior is observed, suggesting strong coupling between individual polar layers. This is consistent with recent second-harmonic generation results [A.Q. Jiang *et al.*, *J. Appl. Phys.* **93**, 1180 (2003)] of ordering in $SrTiO_3/BaTiO_3$ superlattices. However, both constituents of the structures investigated here are paraelectric. Strain-induced room-temperature ferroelectricity in $SrTiO_3$ and distance-dependent coupling between these layers are proposed as mechanisms leading to the observed behavior.

The periodic stacking of epitaxial perovskite films, and thus the formation of artificial superlattice structures, allows us to intimately couple dissimilar materials and to observe emerging physical properties that are not necessarily a simple combination of those found in the constituent materials. Motivated both by the technological interest in ferroelectrics for device applications and the quite good understanding of these materials' intriguing properties, structures consisting of paraelectric and ferroelectric layers have received considerable attention. Here we report on the observation of ferroelectric and antiferroelectric properties in structures consisting entirely of paraelectric constituents, and show that the data are compatible with an interpretation of strain-induced ferroelectricity at room temperature and spacing-dependent coupling between such layer.

Periodic heterostructures consisting of paraelectric and ferroelectric perovskite titanate or niobate layers have been studied in detail before [1-5]. In the case of $KTaO_3/KNbO_3$ superlattices, for example, it was observed that below a critical layer spacing, the structural phase transition occurs at the same temperature as that of the alloy [6,7], but the local structure remains distinctively different from that of the solid-solution [8]. Furthermore, dielectric measurements show evidence of anti-ferroelectricity in 1/1 superlattices [9].

$SrTiO_3$ is typically described as a quantum paraelectric, i.e. a material in which ferroelectricity would occur at low temperature if it were not for the quantum fluctuations. The lattice is easily be distorted by impurities (such as Ca or Ba on the Sr-site), leading to local polar clusters or ferroelectric states. In addition, the dielectric constant is strongly pressure-sensitive [10-12], and a transition to a ferroelectric state at low temperature can be induced by uniaxial stress [13,14].

More recently, a careful treatment of $SrTiO_3$ films in the presence of misfit strains (resulting from film/substrate interactions) in the framework of the Landau-Ginsburg-Devonshire theory has led to a rather complete description of ferroelectricity in these layers [15]. Room-temperature ferroelectricity is predicted for large strains of about 0.015 but has not been confirmed experimentally.

Epitaxial superlattices provide the ideal platform to probe the effects of such large misfit strains on $SrTiO_3$. According to a recent report [16], second-harmonic generation data show that $SrTiO_3$ exhibits a polar state at room temperature in $SrTiO_3/BaTiO_3$ superlattices if the $SrTiO_3$ layer thickness falls below 30 unit cells (i.e. 30/30 superlattices). Interestingly, "antidipole patterns" are observed in superlattices with unit cells below 10/10, corresponding to anti-ferroelectric ordering of the structure.

Dielectric measurements can shed light onto the properties of such materials; however, these experiments probe the entire structure, and the properties of a material that is not ferroelectric in its relaxed state may be partially or completely masked by the ferroelectric constituent of the superlattice. In order to circumvent this difficulty, we performed dielectric measurements on superlattices of SrTiO$_3$ and BaZrO$_3$. Both of these materials are cubic (Pm$\bar{3}$m) and paraelectric at room temperature. Our data are consistent with an interpretation (motivated by the above-mentioned study of the related SrTiO$_3$/BaTiO$_3$ structures) of a polar state in SrTiO$_3$ (and/or in BaZrO$_3$) having long-range (inter-layer) order of ferroelectric or anti-ferroelectric nature depending on the stacking periodicity (ferroelectricity for (20/20) and (38/38) structures, anti-ferroelectricity for (15/15) and (7/7) structures).

Epitaxial films and superlattices were grown onto LaAlO$_3$ substrates by pulsed laser deposition (PLD) under standard growth conditions (KrF radiation, 2 J/cm$^2$ at 10 Hz, 200 mTorr oxygen background, and substrates mounted with silver paint to a plate kept at 780 °C). Single-phase ceramic targets were used, and all films were grown with a total thickness of 500 nm.

The superlattices were first characterized by x-ray diffraction. As shown in Fig. 1, satellite peaks are clearly observed in normal θ-2θ scans (using a Rigaku 2-circle diffractometer) and allow us to determine the periodicity of the structure accurately. Profilometry measurements on single BaZrO$_3$ and SrTiO$_3$ films indicate that the growth rates are identical (± 5%) for the two materials; thus the thickness of each constituent layer is half of the superlattice periodicity.

For the samples with larger periodicity, x-ray diffraction allowed us to determine the in-plane and out-of-plane lattice parameters of both constituents. Out-of-plane measurements were made with a four-circle diffractometer, using Cu Kα radiation, and a sagitally-focusing focusing graphite monochromator. Reflections were broad, indicating that the layers were incoherent. Measuring the Bragg angle for the (110), (200), (220), (101), (211), (112), and (202) peaks on a (20/20) superlattice shows that the films are strongly strained. Table I shows the results for each material's lattice parameter.

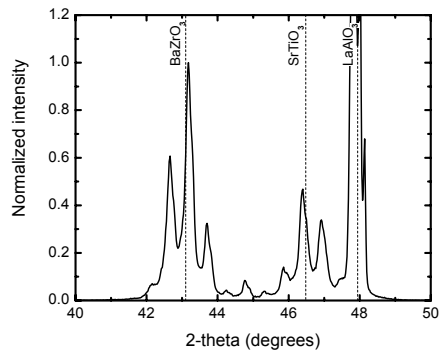

FIG. 1. Normal x-ray θ-2θ scan for a superlattice consisting of BaZrO$_3$ and SrTiO$_3$ layers on a LaAlO$_3$ substrates. From the spacing of the satellite peaks, the periodicity of the structure ($\Lambda$ = 180 Å) can be determined.

Obviously, the strain in these films results not from clamping to the substrate but from the interaction between the SrTiO$_3$ and the BaZrO$_3$: For SrTiO$_3$, *c/a < 1*, indicating tensile stress resulting from the larger BaZrO$_3$ rather than compression from the smaller LaAlO$_3$. The (110), (200), and (220) peaks are measured in a glancing-incidence geometry. They are therefore not affected by broadening from finite layer thickness, leaving peak broadening due to finite in-plane grain size and stress inhomogeneity. Noting that the (220) peak is twice as broad as the (110), we can eliminate small grain size effects as a possible cause of line broadening, leaving a stress inhomogeneity of 1.8% FWHM. This indicates that the material near the interface may be under more strain than the middle of each layer.

|  | Bulk $a_0$ (Å) | Film in-plane $a$ (Å) | Film normal $c$ (Å) | Intrinsic (cubic) film lattice para. $\hat{a}$ | In-plane strain $\varepsilon_{xx} = (a - \hat{a})/\hat{a}$ | Normal strain $\varepsilon_{zz} = (c - \hat{a})/\hat{a}$ |
|---|---|---|---|---|---|---|
| SrTiO$_3$ | 3.905 | 3.955 | 3.918 | 3.943 (ν=.232) <br> 3.932 (ν=.5) | 5.9·10$^{-3}$ (ν=.232) <br> 3.1·10$^{-3}$ (ν=.5) | -3.5·10$^{-3}$ (ν=.232) <br> -6.2·10$^{-3}$ (ν=.5) |
| BaZrO$_3$ | 4.193 | 4.161 | 4.212 | 4.178 (ν=.232) <br> 4.193 (ν=.5) | -7.6·10$^{-3}$ (ν=.232) <br> -4.1·10$^{-3}$ (ν=.5) | 4.6·10$^{-3}$ (ν=.232) <br> 8.2·10$^{-3}$ (ν=.5) |
| LaAlO$_3$ | 3.788 |  |  |  |  |  |

**Table I.** Observed lattice parameters of SrTiO$_3$ and BaZrO$_3$ in a (20/20) superlattice on LaAlO$_3$. The intrinsic film lattice parameter (relaxed cell volume) and the strain values are given for the bulk SrTiO$_3$ Poisson's ratio (ν = 0.232) and for the limiting case of an incompressible solid (ν = 0.5).

For the samples with smaller periodicity (7/7 and 15/15), it was impossible to determine the in-plane and out-of-plane lattice parameters of the constituents independently. We can, however, safely assume that strain values obtained for the 20/20 structure represents a lower limit for the actual values present in the 7/7 and 15/15 structures.

From the data in Table I, the strain in each of the layers can be obtained. First, the intrinsic relaxed (cubic) lattice parameter of the $SrTiO_3$ and $BaZrO_3$ layers is calculated as

$$\hat{a} = \frac{(1-\nu)c + 2\nu a}{1+\nu}$$

where $\nu$ is Poisson's ratio. For an incompressible solid, $\nu = 0.5$; for bulk $SrTiO_3$, $\nu = 0.232$ [17]. The Poisson's ratios for thin films of $SrTiO_3$ and $BaZrO_3$ are not known, however, the in-plane strain, $\varepsilon_{xx} = (a - \hat{a})/\hat{a}$, and the out-of-plane (or normal) strain, $\varepsilon_{zz} = (c - \hat{a})/\hat{a}$, can be calculated for different values of $\nu$ and are found to be in the range of a fraction of a percent, as shown in Table I.

Not surprisingly, the films' intrinsic lattice parameter differs from that of the bulk value, as often observed for PLD-grown films due to a high density of point defects.

For dielectric measurements, interdigital (co-planar) Au/Cr electrode structures were deposited onto the film surfaces, and the measurements were performed at 10 kHz. The finger spacing of these electrodes was 10 μm; thus a bias voltage of 1V corresponds to an electric field of 1 kV/cm. Measurements were performed either at room temperature or by immersing the entire structure in liquid nitrogen.

Figure 2 shows room-temperature dielectric data for various structures. The data in (a) and (b) are obtained for reference films of $BaZrO_3$ and $SrTiO_3$. As expected, little variation is observed in the capacitance (and thus the dielectric constant) as function of applied dc voltage. The superlattices, in contrast, exhibit an interesting electric-field dependence. Clearly, their behavior cannot be understood as a simple combination of the $BaZrO_3$ and $SrTiO_3$ properties, despite the fact that the in-plane measurement geometry would allow us to view the sample as a parallel-combination of individual $BaZrO_3$ and $SrTiO_3$ layers. For the samples with the smallest periodicity (but with the same total sample thickness of 500 nm), the capacitance curves exhibit a local minimum at 0 kV/cm, and two symmetrically placed maxima at a value that depends on the periodicity. For larger structures, such as shown in Fig. 2(e) (and also in Fig. 3, see below), a conventional (hysteretic) butterfly-loop is observed. This type of curve is typically associated with ferroelectric structures. Note that here, however, neither of the two constituent materials of the superlattice is independently ferroelectric.

Figure 3 shows the capacitance vs. voltage data for a 500 nm thick film consisting of 16 pairs of $SrTiO_3$ and $BaZrO_3$, each layer being approximately 38 unit cells thick. Ferroelectric-type behavior is observed both at room temperature and at 77 K.

The temperature-dependence of the out-of-plane film lattice parameter was recorded from room temperature to 650 °C by x-ray diffraction (data not shown). Quite surprisingly, no anomaly is observed, i.e. the thermal expansion is linear over the entire range with a scatter of about $4 \cdot 10^{-5}$. Thus, if a structural transition were present in this temperature range (similar to our earlier work on $KTaO_3/KNbO_3$ [6] ), it would clearly be detected. It is important to point out, however, that the $(c-a)/a$ distortions in the $KTaO_3/KNbO_3$ change by only about 30% at the transition, and that the distortion observed here ($(c-a)/a \approx -9 \cdot 10^{-3}$) is comparable in magnitude to that observed in $KTaO_3/KNbO_3$ even in the high-temperature phase ($(c-a)/a \approx 6 \cdot 10^{-3}$).

Capacitance-vs.-voltage (C(V)) curves are, by definition, the derivative (with respect to the electric field) of polarization-vs.-field (P(E)) loops (in our case with the dc field swept at a very low rate of about $10^{-2}$

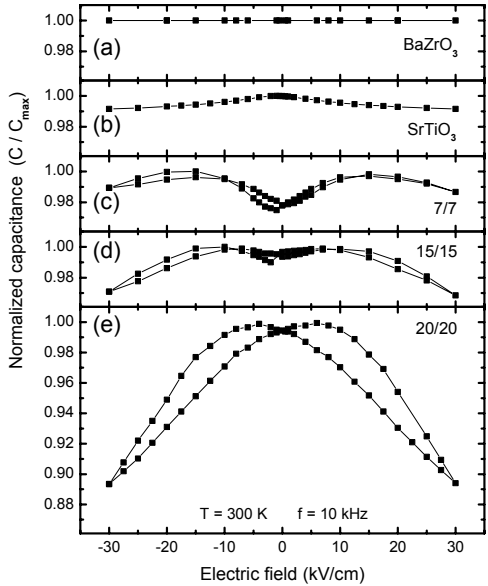

FIG. 2. Capacitance-voltage curves obtained using coplanar (inter-digital) electrodes on various films on $LaAlO_3$ substrates at room temperature. (a) and (b) show the data for single reference films of $BaZrO_3$ and $SrTiO_3$, respectively. In (c), (d), and (e), the data is presented for superlattices in which the $BaZrO_3$ and $SrTiO_3$ are 7, 15, and 20 unit cells thick, respectively. Total film thickness for all samples is 500 nm.

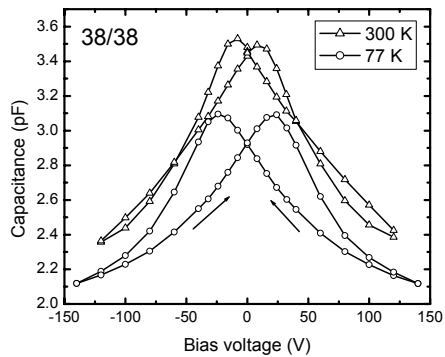

FIG. 3. Capacitance-voltage curves for a superlattice consisting of a periodic stacking of 38 unit-cell thick layers of $BaZrO_3$ and $SrTiO_3$, measured at room temperature and in liquid nitrogen.

Hz, but the derivative taken at 10 kHz). Therefore, a single hysteresis loop (in P(E)) yields the traces as observed in our 20/20 and 38/38 structures. Anti-ferroelectricity, or anti-ferroelectric dipole ordering between polar layers [18], is reflected by double hysteresis-loops in P(E), and thus by two maxima in the C(V) curves (with a minimum at V=0), as observed in our 7/7 and 15/15 structures.

While our results by themselves may be insufficient to demonstrate ferroelectricity or anti-ferroelectricity (for example, the electrode geometry is inappropriate for recording polarization hysteresis loops), the observation of similar dipole patterns in $BaTiO_3/SrTiO_3$ and $KTaO_3/KNbO_3$ superlattices strongly support such an interpretation. Our data is thus fully consistent with a picture of polar (ferroelectric) ordering within the $SrTiO_3$ and/or the $BaZrO_3$ layers, and a long-range (inter-layer) polar order of anti-ferroelectric nature for spacing and thickness of 15 or fewer unit cells. For larger layer spacing and thickness, these ferroelectric slabs either show no long-range correlation or order ferroelectrically.

Anti-ferroelectric ordering between individual ferroelectric layers in a superlattice may be understood from energy arguments. In fact, the long-range nature of electric dipolar interaction leads to a large energy associated with long-range ferroelectric ordering. In ferroelectric crystals, where the ordering is mediated via elastic (short-range) interactions, domain formation reduces the energy associated with the dipolar long-range interactions. The energy associated with the formation of domain walls determines the size of domains. Just as in the case of ferromagnetic films [19], the equilibrium domain size in a ferroelectric layer scales as $D \propto \sqrt{t}$ and thus becomes larger than the film thickness below a critical value [6]. At this point, the equilibrium configuration is no longer determined by the properties of the individual layers in a superlattice but by the total material within a domain.

If the material has a structure that elastically allows for the formation of anti-ferroelectric ordering – as is the case in superlattices – anti-ferroelectric ordering will be favorable for certain values of elastic and dipolar interaction strengths.

The question remains, however, where the polarization resides. There are three obvious possibilities: either $SrTiO_3$ or $BaZrO_3$ could be ferroelectric under these strained conditions, or the interface (and possibly an inter-diffusion layer) could be responsible for the behavior. In fact, Ba-diffusion into $SrTiO_3$ would render this material ferroelectric, but much more than 50% Ba needs to be substituted for Sr in order to achieve room-temperature ferroelectricity, even if it is assumed that Zr doesn't substitute for Ti. More importantly, however, in our previous work on superlattices of $SrTiO_3$ and $BaTiO_3$ [20], where the same type of diffusion would be expected, we have observed distinctively different behavior between 4/4 superlattices and alloys. This indicates that under the growth conditions used here, interdiffusion – if it occurs – is restricted to a length scale of less than 4 unit cells. Ferroelectric-type behavior in the present $SrTiO_3/BaZrO_3$ superlattices, however, is clearly apparent in structures with periodicities ten times larger. We can therefore rule out interdiffusion as the dominating factor.

Consistent with the above-mentioned work on $LaAlO_3/SrTiO_3$ structures, strain-induced room-temperature ferroelectricity in $SrTiO_3$ appears to be most likely. In fact, the earlier work on single-crystal materials [13,14] has shown that low-temperature ferroelectricity occurs in uniaxially stressed $SrTiO_3$. In this situation, the lattice distortion is of the same type as in this work, i.e. the unit cell dimensions are reduced in one direction and extended in the other two. Data for the actual distortions in the bulk experiments are not available, but an estimation can be made using published values of the elastic moduli ($c_{11} = 2.95 \cdot 10^{11}$ N/m$^2$ and $c_{12} = 1.15 \cdot 10^{11}$ N/m$^2$ at 30 K [21]. For the highest stress investigated, $\sigma = 4 \cdot 10^8$ N/m$^2$ (for which a ferroelectric transition temperature $T_c \approx 30$ K was observed), these values yield $\Delta a/a \approx 4.9 \cdot 10^{-4}$, and $\Delta c/c \approx -1.7 \cdot 10^{-3}$. The fact that the distortions in our films are significantly larger supports the interpretation of a strain-induced ferroelectric phase in $SrTiO_3$ at room temperature. It appears, however, that much less strain is required in this symmetric environment to drive $SrTiO_3$ into a ferroelectric state than the value of $1.5 \cdot 10^{-2}$ calculated for $SrTiO_3$ films with a free surface [15].

To conclude, $SrTiO_3/BaZrO_3$ superlattices exhibit dielectric properties that are very different from what would be expected based on reference data on $SrTiO_3$ and $BaZrO_3$ films. Strain-induced room-temperature ferroelectricity in SrTiO$_3$ and spacing-dependent coupling between the layers are proposed as mechanisms leading to this behavior. For large superlattice periodicities (20/20 and 38/38 structures), these ferroelectric layers appear to act as independent ferroelectric slabs or exhibit long-range ferroelectric ordering. For smaller periodicities (7/7 and 15/15), anti-ferroelectric coupling between these polar layers is observed, consistent with energy considerations for long-range electrostatic interactions and domain formation.